\documentclass[twocolumn,preprintnumbers,amsmath,amssymb]{revtex4}
\usepackage{graphicx}
\usepackage{dcolumn}
\usepackage{bm}

\begin{document}
\title{Interference-assisted squeezing in fluorescence radiation}
\author{R. Arun}
\affiliation{Department of Physics, Faculty of Engineering $\&$ Technology, SRM University,
Kattankulathur 603 203, Tamilnadu, India.}


\begin{abstract}
The squeezing spectrum of the resonance fluorescence is studied for a coherently driven
four-level atom in the Y-type configuration. It is found that the squeezing properties
of the fluorescence radiation are modified significantly when quantum interference of the
spontaneous decays channels is included. We show a considerable enhancement of
steady-state squeezing in spectral components for strong and off-resonant driving fields.
The squeezing may be increased in both the inner and outer sidebands of the spectrum depending
upon the choice of parameters. We also show that the interference can degrade the spectral
squeezing by increasing the decay rates of atomic transitions. An analytical description using
dressed states is provided to explain the numerical results. 
\end{abstract}

\maketitle

\newpage
\section{Introduction}
Squeezing of the radiation field is one of the distinct features of the quantum theory
of light \cite{review,mandel}. Squeezed states of light, being nonclassical in origin \cite{patnaik},
have a reduced variance in one of the quadrature components of the electric field.
It is well known that the resonance fluorescence from a driven atomic system can serve as a
source of squeezed radiation. Theoretical investigations on two-level and three-level atoms
demonstrated squeezing either in the total variance of phase quadratures or in the frequency
(spectral) components of the fluorescence radiation
\cite{zoller,collet,swain,expt,three,dalton,gao1,gao2}. For a driven two-level atom, Walls
and Z\"{o}ller first predicted that squeezing can occur in the in-phase/out-of-phase
quadrature component of the fluorescent light [4a]. The noise spectrum was
shown to exhibit single- and two-mode squeezing in the weak- and  strong-excitation regimes
\cite{collet,swain}. Some experimental evidences of squeezing have also been reported in
the phase-dependent spectra of two-level atoms \cite{expt}. Unlike in two-level systems,
two-photon coherences play a significant role in the dynamics of three-level atoms driven by
coherent fields \cite{three,dalton}. Dalton {\it et al.} examined the role of atomic
coherences and studied the maximum squeezing that can be obtained in the fluorescence
from  three-level systems \cite{dalton}. A detailed study by Gao {\it et al.}
\cite{gao1,gao2} has shown that ultranarrow squeezing peaks may appear in the spectrum of
driven three-level atoms in $\Xi$- and V-type configurations. However, the role of two-photon
coherence is seen to destroy the spectral-component squeezing in the fluorescent field \cite{gao1}.

One of the interesting developments in the study of resonance fluorescence is
the possibility of modifying spectral properties of the atoms via quantum interferences
in spontaneous decay channels. The interference in spontaneous emission occurs when the atomic
transitions are coupled by same vacuum modes. The early work of Agarwal on this subject demonstrated
population trapping and generation of quantum coherence between the excited states in a V-type atom \cite{gsa}.
Since the fluorescence properties of a  driven atomic system result from its spontaneous emission, studying
the influence of interference in such processes  has become an important topic of research
\cite{cardi,li1,arun,anton1,li2,anton2}. Much attention has been paid to study the  fluorescence spectrum of
driven atoms \cite{cardi,li1,arun}. All these theoretical studies assume non-orthogonal dipole moments of the
atomic transitions for the interference to exist in decay processes \cite{gsa}. However, in real atomic systems,
it is difficult to meet this condition. Different schemes were later proposed to bypass the condition of non-orthogonal
dipole moments \cite{preselect,qdot,carmichael}. Experimentally, coherence between the ground states arising from
spontaneous emissions has been reported using electron spin polarization states in quantum dots \cite{qdot} and zeeman
sub-levels in atomic systems \cite{carmichael}.

The squeezing characteristics of the emitted fluorescent light was also discussed extensively \cite{anton1,li2,anton2}.
In a three-level V-type atom interacting with a coherent field, the interference is shown to enhance the
spectral component squeezing in the presence of standard/squeezed vacuum \cite{anton1}.
Li {\it et al.} studied the squeezing spectrum of a four-level atom in the
$\Lambda$-type scheme and showed that unusual squeezing properties appear due to
interference if the fluorescent field is detected on the slow decaying transition
\cite{li2}. Further, Gonzalo {\it et al.} have examined a driven  three-level atom of the $\Lambda$ configuration
with particular attention to the squeezing in spectral components \cite{anton2}. Recently, the effects of spontaneously
generated interferences have been investigated in the context of enhancing self-Kerr nonlinearity \cite{kerr}, soliton
formation \cite{soliton}, and preserving bi-partite entanglement \cite{qubits}.

\begin{figure}[b]
   \centering
    \includegraphics[width=3.6cm]{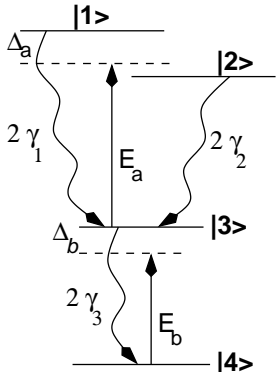}
\caption{The level scheme of the Y-type atom driven by coherent fields.}
\end{figure}

In this paper, we consider a four-level atom in the Y-type configuration interacting with
two coherent fields (as shown in Fig. 1). The excited atomic states are assumed to be near
degenerate and decay spontaneously via the same vacuum modes to the intermediate state. The atom in
the intermediate state can make spontaneous transitions to the ground state. Since the cascade decays
from the excited atomic states lead to an emission of the same pair of photons,
quantum interference exists in decay processes. The role of the interference was investigated
in the fluorescence spectrum of this system in Ref. \cite{arun}. In the present work, we
study the squeezing spectrum and examine the interference effects on squeezing properties of
the fluorescence fields.

The paper is arranged as follows. In Sec. II, we present the atomic density matrix equations,
describing the interaction of a Y-type atom with two coherent fields, when the presence of
quantum interference in decay channels is included. The formula for the squeezing spectrum
is then derived using atomic correlation operators in Sec III. In Sec. IV, we analyze
the numerical results of the squeezing spectrum and identify the origin of interference effects
using the dressed-state picture. Finally, the main results are summarized in Sec. V.

\section{driven Y-type atomic system and its density matrix equations}
We consider a four-level atom in the Y-type configuration as shown in Fig. 1. In this scheme,
the atom has two closely lying excited states $|1\rangle$ and $|2\rangle$ with energy separation
$\hbar W_{12}$. It is assumed that the excited atomic states are coupled by common vacuum modes
to decay spontaneously to the intermediate state $|3\rangle$ with rates  $2 \gamma_1$ and $2 \gamma_2$.
The atom in the intermediate state $|3\rangle$ is further allowed to undergo spontaneous emissions to the
ground state $|4\rangle$ with decay rate $2\gamma_3$. The direct transitions between the excited states
$|1\rangle \rightarrow |2\rangle$ and that between the excited and ground states
$|1\rangle, |2\rangle \rightarrow |4\rangle$ of the atom are forbidden in the dipole
approximation. We assume that the transition frequencies $(\omega_{13},\omega_{23})$ of the upper transitions
differ widely from that of the lower transition $(\omega_{34})$. This leads to a situation in which the vacuum modes
coupling the upper and lower atomic transitions are totally different. In addition to spontaneous decays, two
coherent fields are applied on the atom as shown schematically in Fig. 1. The upper transitions
$|1\rangle, |2\rangle \leftrightarrow |3\rangle$ are driven by a coherent field of frequency
$\omega_a$ (amplitude $E_a$) and another field of frequency $\omega_b$ (amplitude $E_b$) couples the
lower transition $|3\rangle \leftrightarrow |4\rangle$. The Rabi frequencies of the atom-field interaction
are denoted as $\Omega_1 = \vec{\mu}_{13}.\vec{E}_a/\hbar$, $\Omega_2 = \vec{\mu}_{23}.\vec{E}_a/\hbar$, and
$\Omega_3 = \vec{\mu}_{34}.\vec{E}_b/\hbar$ with $\vec{\mu}_{mn}$ being the dipole moment of the atomic transition
from $|m\rangle$ to $|n\rangle$.

The system is studied in the interaction picture using time independent Hamiltonian
\begin{eqnarray}
H_I &=& \hbar (\Delta_a + \Delta_b) A_{11} + \hbar (\Delta_a + \Delta_b - W_{12}) A_{22}
\nonumber \\
&& + \hbar \Delta_b A_{33} -\hbar (\Omega_1 A_{13} + \Omega_2 A_{23} + \hbox{H.c.}) \nonumber \\
&& - \hbar (\Omega_3 A_{34} + \hbox{H.c.}).  \label{ham}
\end{eqnarray}
Here, the operators $A_{mn} = |m\rangle\langle n|$ represent the atomic population operators for
$m=n$ and transition operators for $m \neq n$. To include decay processes in the dynamics, we use
the master equation framework. With the inclusion of the decay terms, the time evolution of the density
matrix elements in the interaction picture obeys \cite{arun}
\begin{equation}
\dot{\rho}_{11} = -2 \gamma_1 \rho_{11} + i \Omega_1 (\rho_{31} - \rho_{13})
- p \sqrt{\gamma_1 \gamma_2} (\rho_{12} + \rho_{21}), \label{rho1}
\end{equation}
\begin{equation}
\dot{\rho}_{22} = -2 \gamma_2 \rho_{22} + i \Omega_2 (\rho_{32} - \rho_{23})
- p \sqrt{\gamma_1 \gamma_2} (\rho_{12} + \rho_{21}),
\end{equation}
\begin{eqnarray}
\dot{\rho}_{33} =&& 2 \gamma_1 \rho_{11} + 2 \gamma_2 \rho_{22} - 2 \gamma_3 \rho_{33}
+ i \Omega_1 (\rho_{13} - \rho_{31})  \nonumber  \\
&& + i \Omega_2 (\rho_{23} - \rho_{32}) + i \Omega_3 (\rho_{43} - \rho_{34}) \nonumber  \\
&& + 2 p \sqrt{\gamma_1 \gamma_2} (\rho_{12} + \rho_{21}),
\end{eqnarray}
\begin{eqnarray}
\dot{\rho}_{12} = && - (\gamma_1 + \gamma_2 + i W_{12}) \rho_{12} + i \Omega_1 \rho_{32}
- i \Omega_2 \rho_{13}  \nonumber \\
&& - p \sqrt{\gamma_1 \gamma_2} (\rho_{11} + \rho_{22}),
\end{eqnarray}
\begin{eqnarray}
\dot{\rho}_{13} = && - (\gamma_1 + \gamma_3 + i \Delta_a) \rho_{13} + i \Omega_1
(\rho_{33} - \rho_{11}) - i \Omega_2 \rho_{12} \nonumber \\
&& - i \Omega_3 \rho_{14} - p \sqrt{\gamma_1 \gamma_2}~ \rho_{23},
\end{eqnarray}
\begin{eqnarray}
\dot{\rho}_{23} =&& - [\gamma_2 + \gamma_3 + i (\Delta_a - W_{12})] \rho_{23} +
i \Omega_2 (\rho_{33} - \rho_{22})  \nonumber \\
&& - i \Omega_1 \rho_{21} - i \Omega_3 \rho_{24} - p \sqrt{\gamma_1 \gamma_2}~ \rho_{13},
\end{eqnarray}
\begin{eqnarray}
\dot{\rho}_{34} = && - (\gamma_3 + i \Delta_b) \rho_{34} + i \Omega_3 (\rho_{44} - \rho_{33})
 + i \Omega_1 \rho_{14} \nonumber \\
&& + i \Omega_2 \rho_{24},
\end{eqnarray}
\begin{eqnarray}
\dot{\rho}_{14} = && - [\gamma_1 + i (\Delta_a + \Delta_b)] \rho_{14} + i \Omega_1 \rho_{34}
- i \Omega_3 \rho_{13}  \nonumber \\
&& - p \sqrt{\gamma_1 \gamma_2}~ \rho_{24},
\end{eqnarray}
\begin{eqnarray}
\dot{\rho}_{24} = && - [\gamma_2 + i (\Delta_a + \Delta_b - W_{12})] \rho_{24} + i \Omega_2
\rho_{34} - i \Omega_3 \rho_{23}  \nonumber \\
&& - p \sqrt{\gamma_1 \gamma_2}~ \rho_{14}.
\label{rho9}
\end{eqnarray}
Here, $\Delta_a = \omega_{13} - \omega_a$ corresponds to the detuning between the atomic
frequency $(\omega_{13})$ of the $|1\rangle \rightarrow |3\rangle$ transition
and the frequency of the applied field $E_a$. Similarly, $\Delta_b = \omega_{34}
- \omega_b$ denotes the detuning of the field acting on the lower transition.
In writing Eqs. (\ref{rho1})-(\ref{rho9}), we have assumed that the trace condition
$\rho_{11}+\rho_{22} +\rho_{33}+\rho_{44}=1$ is satisfied. The cross-coupling term
$p \equiv \vec{\mu}_{13}.\vec{\mu}_{23}/|\vec{\mu}_{13}|
|\vec{\mu}_{23}|$ is referred to as the interference parameter. This term arises due to the
quantum interference in spontaneous emission pathways. The effect of interference is to couple
the populations and coherences as seen in Eqs. (\ref{rho1})-(\ref{rho9}). It reflects the fact
that population can be transferred between the excited states by the vacuum field.
When $p = \pm 1$, the decays from the excited states $|1\rangle$ and $|2\rangle$ are
coupled and the interference effects are maximal. If the atomic dipole moments are orthogonal $(p=0)$,
there is no interference effect in spontaneous emission.

For convenience in the calculation of the squeezing spectrum, we rewrite the density matrix equations
(\ref{rho1})-(\ref{rho9}) in a more compact matrix form by the definition
\begin{eqnarray}
\hat{\Psi} = && \left(\rho_{11},\rho_{22},\rho_{33},\rho_{12},\rho_{13},\rho_{23},\right.
\nonumber \\
&& \left. \times \rho_{14}, \rho_{24},\rho_{34},\rho_{21},\rho_{31},\rho_{32},\rho_{41},\rho_{42},
\rho_{43}\right)^{T}.
\label{psidef}
\end{eqnarray}
Substituting Eq. (\ref{psidef}) into Eqs. (\ref{rho1})-(\ref{rho9}), we get the
matrix equation for the variables $\hat{\Psi}_j(t)$
\begin{equation}
\frac{d}{dt} \hat{\Psi} = \hat{L} \hat{\Psi} + \hat{I}, \label{matrix}
\end{equation}
where $\hat{\Psi}_j$ is the $j$-th component of the column vector $\hat{\Psi}$ and the
inhomogeneous term $\hat{I}$ is also a column vector with non-zero components
\begin{equation}
\hat{I}_9 = i \Omega_3,~~~~~\hat{I}_{15} = -i \Omega_3.
\end{equation}
In Eq. (\ref{matrix}), $\hat{L}$ is a 15$\times$15 matrix whose elements are
time independent and can be found explicitly from Eqs. (\ref{rho1})-(\ref{rho9}).
The steady-state solutions of the density matrix elements can be obtained by setting the
time derivative equal to zero in Eq. (\ref{matrix}):
\begin{equation}
\hat{\Psi}(\infty) = - \hat{L}^{-1} \hat{I}. \label{steady}
\end{equation}

\section{phase dependent fluorescence spectrum}
We now proceed to the study of the squeezing spectra of the driven atom. Since the atom
is driven by two coherent fields, each field induces its own atomic dipole moment which
then generates a scattered field. However, the fields scattered by the upper and lower
transitions in the atom will have no correlations because the applied fields $(E_a,E_b)$
are of quite different carrier frequencies $(\omega_a,\omega_b)$. Assuming that the point
of observation lies perpendicular to the atomic dipole moments, the positive-frequency part
of the fluorescent fields in the radiation zone can be written as
\begin{eqnarray}
\vec{E}^{(+)}_a(t) &=& f(r)[\vec{\mu}_{13} A_{31}(\hat{t}) + \vec{\mu}_{23}
A_{32}(\hat{t})]\exp(-i\omega_a\hat{t}), \nonumber \\
\vec{E}^{(+)}_b(t) &=& g(r) \vec{\mu}_{34} A_{43}(\hat{t})
\exp(-i\omega_b\hat{t}), \label{electric}
\end{eqnarray}
where $\hat{t} = t - r/c$, $f(r) = \omega_{13}^2/c^2r$, $g(r) = \omega_{34}^2/c^2r$, and $r$ is
the distance of the detector from the atom. The index $a$ $(b)$ in Eq. (\ref{electric}) refers
to the fluorescent light of central frequency $\omega_a$ $(\omega_b)$. In squeezing measurements,
the two-time expectation value of a particular quadrature component of the electric field is
the quantity of interest. We consider squeezing in the fluorescent light exclusively emitted by
the upper and lower transitions in the atom.  The slowly varying quadrature components with
phase $(\theta)$ are defined as
\begin{eqnarray}
\vec{E}_a(\theta,t) &=& \vec{E}^{(+)}_a(t) e^{i(\omega_at+\theta)} + \vec{E}^{(-)}_a(t)
e^{-i(\omega_at+\theta)},\nonumber \\
\vec{E}_b(\theta,t) &=& \vec{E}^{(+)}_b(t) e^{i(\omega_bt+\theta)} + \vec{E}^{(-)}_b(t)
e^{-i(\omega_bt+\theta)}.
\end{eqnarray}
The spectrum of squeezing is defined by the Fourier transformation of the normal and time-ordered
correlation of the quadrature component $\vec{E}(\theta,t)$:
\begin{eqnarray}
S_a(\omega,\theta) &=& \frac{1}{2\pi} \int_{-\infty}^{\infty} \hat{T}
\langle:\vec{E}_a(\theta,t),\vec{E}_a(\theta,t+\tau):\rangle e^{i\omega\tau}d\tau,
\nonumber \\
S_b(\omega,\theta) &=& \frac{1}{2\pi} \int_{-\infty}^{\infty} \hat{T}
\langle:\vec{E}_b(\theta,t),\vec{E}_b(\theta,t+\tau):\rangle e^{i\omega\tau}d\tau, \nonumber\\
\label{spect}
\end{eqnarray}
where $\langle\vec{A},\vec{B}\rangle = \langle\vec{A}.\vec{B}\rangle - \langle\vec{A}\rangle .
\langle\vec{B}\rangle$, $\hat{T}$ represents the time ordering operator, and
the steady-state limit $(t \rightarrow \infty)$ is considered.

The calculations of the correlation functions in Eq. (\ref{spect}) can be carried out easily
with the help of the quantum regression theorem and the density matrix equations (\ref{matrix}).
For this purpose, we introduce column vectors of two-time averages
\begin{eqnarray}
&&\hat{U}^{mn}(t,\tau) =     \nonumber \\
&& \left[\langle \Delta A_{11}(t+\tau) \Delta A_{mn}(t)\rangle,
\langle \Delta A_{22}(t+\tau) \Delta A_{mn}(t)\rangle, \right. \nonumber \\
&& \langle \Delta A_{33}(t+\tau) \Delta A_{mn}(t)\rangle, \langle \Delta A_{21}(t+\tau)
\Delta A_{mn}(t)\rangle,  \nonumber \\
&& \langle \Delta A_{31}(t+\tau) \Delta A_{mn}(t)\rangle, \langle \Delta A_{32}(t+\tau)
\Delta A_{mn}(t)\rangle, \nonumber \\
&& \langle \Delta A_{41} (t+\tau) \Delta A_{mn}(t)\rangle, \langle \Delta A_{42}(t+\tau)
\Delta A_{mn}(t)\rangle, \nonumber \\
&& \langle \Delta A_{43}(t+\tau) \Delta A_{mn}(t)\rangle, \langle \Delta A_{12}(t+\tau)
\Delta A_{mn}(t)\rangle, \nonumber \\
&& \langle \Delta A_{13}(t+\tau) \Delta A_{mn}(t)\rangle, \langle \Delta A_{23}(t+\tau)
\Delta A_{mn}(t)\rangle, \nonumber \\
&& \langle \Delta A_{14}(t+\tau) \Delta A_{mn}(t)\rangle, \langle \Delta A_{24}(t+\tau)
\Delta A_{mn}(t)\rangle, \nonumber \\
&& \left. \langle \Delta A_{34}(t+\tau) \Delta A_{mn}(t)\rangle \right]^{T},
m,n = 1,2,3,4. \label{corre}
\end{eqnarray}
Here, $\Delta A_{mn}(t) = A_{mn}(t) - \langle A_{mn}(\infty) \rangle$ are the deviations of the
atomic operators from its steady state values Eq. (\ref{steady}). According to the quantum
regression theorem \cite{lax}, the column vectors (\ref{corre}) satisfy
\begin{equation}
\frac{d}{d\tau} \hat{U}^{mn}(t,\tau) = \hat{L} \hat{U}^{mn}(t,\tau).
\end{equation}
Now, following the procedure explained in Refs. \cite{anton1,li2,anton2} for the time ordering of
operators in Eq. (\ref{spect}), the squeezing spectrum can be obtained as
\begin{eqnarray}
S_a(\omega,\theta) &=&  \frac{f(r)^2}{\pi} \hbox{Re}\Bigg\{ \sum_{k = 1}^{15}
\lim_{t \rightarrow \infty} \left[ \hat{M}_{5,k} \left( {|\vec{\mu}_{13}|}^2 \hat{U}^{31}_k(t,0)\right.
\right. \nonumber \\
&& \left. + \vec{\mu}_{13}.\vec{\mu}_{23} \hat{U}^{32}_k(t,0)\right)
 + \hat{M}_{6,k} \left({|\vec{\mu}_{23}|}^2 \hat{U}^{32}_k(t,0) \right.\nonumber \\
&& \left. \left. + \vec{\mu}_{13}.\vec{\mu}_{23}\hat{U}^{31}_k(t,0)\right) \right]
e^{i2(\theta + \omega_a r/c)} \label{specta} \\
&& + \sum_{k = 1}^{15} \lim_{t \rightarrow \infty} \left[ \hat{M}_{11,k}
\left( {|\vec{\mu}_{13}|}^2 \hat{U}^{31}_k(t,0) +  \vec{\mu}_{13}.\vec{\mu}_{23} \right. \right. \nonumber \\
&& \left. \times \hat{U}^{32}_k(t,0)
\right) + \hat{M}_{12,k}\left( {|\vec{\mu}_{23}|}^2 \hat{U}^{32}_k(t,0) \right. \nonumber \\
&& + \left. \left. \vec{\mu}_{13}.\vec{\mu}_{23}\hat{U}^{31}_k(t,0)\right) \right]  \Bigg\}, \nonumber
\end{eqnarray}
where $\hat{M}_{j,k}$ denotes the $(j,k)$ element of the matrix $\hat{M} =
[(i \omega - \hat{L})^{-1} + (-i \omega - \hat{L})^{-1}]$. Similarly,
\begin{figure}[t]
   \centering
   \includegraphics[height=6cm]{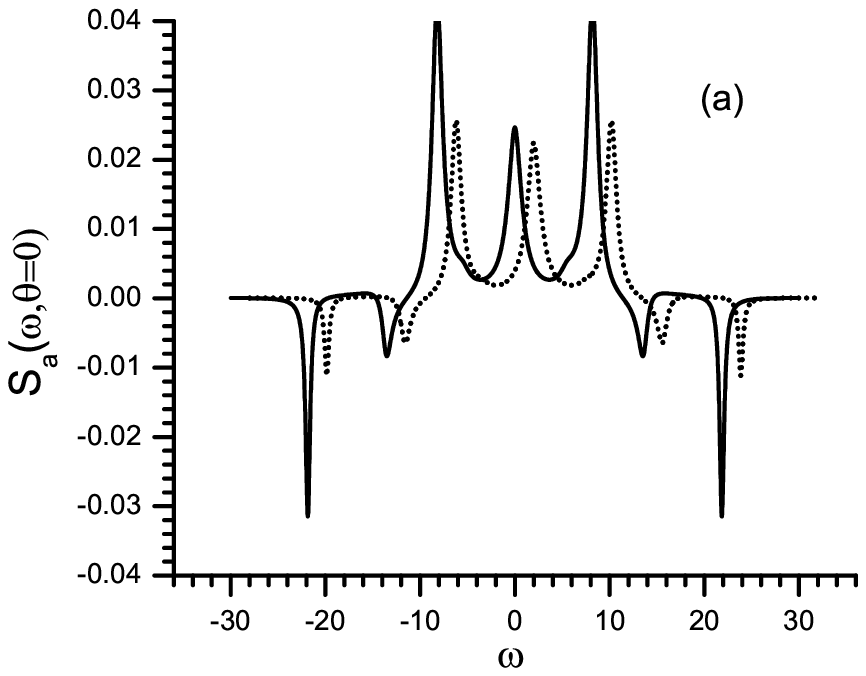}
   \vspace*{-1.2mm}
   \includegraphics[height=6cm]{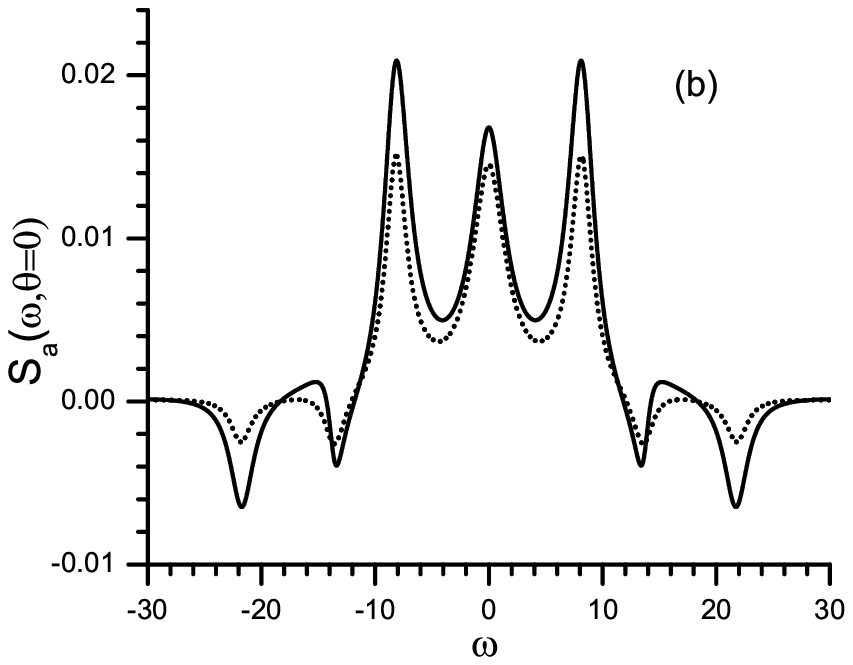}
   \caption{Squeezing spectrum $S_a(\omega,\theta)$ as a function of $\omega$ for the
   parameters $\theta = 0$, $\gamma_3 = 1$, $W_{12} = 10$, $\Delta_a = \Delta_b = 10$, $\Omega_1 = \Omega_2 =
   \Omega_3 = 3$, and $\gamma_1 = \gamma_2 = 0.1$ (a) and $\gamma_1 = \gamma_2 = 1$ (b).
   The solid (dotted) curves are for $p = 1$ $(p = 0)$. For clarity, the dotted curve in (a) has been displaced by
   2 units along the $\omega$-axis.} 	
\end{figure}

\begin{figure}[t]
   \centering
   \includegraphics[height=6cm]{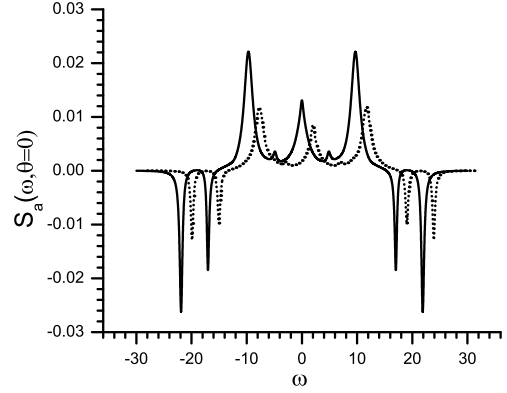}
   \caption{Squeezing spectrum $S_a(\omega,\theta)$ as a function of $\omega$ for the
   parameters $\theta = 0$, $\gamma_3 = 1$, $W_{12} = 5$, $\Delta_a = \Delta_b = 10$, $\Omega_1 = \Omega_2 =
   \Omega_3 = 3$, and $\gamma_1 = \gamma_2 = 0.1$. The solid (dotted) curves are for $p = 1$ $(p = 0)$.
   For clarity, the dotted curve has been displaced by 2 units along the $\omega$-axis.} 	
\end{figure}

\begin{eqnarray}
S_b(\omega,\theta) &=& \frac{{|\vec{\mu}_{34}|}^2 g(r)^2}{\pi} \hbox{Re}\Bigg\{\sum_{k = 1}^{15}
\lim_{t \rightarrow \infty} \left[ \hat{M}_{9,k} \hat{U}^{43}_k(t,0) \right. \nonumber \\
&&~~~\left. \times e^{i2(\theta + \omega_b r/c)} + \hat{M}_{15,k} \hat{U}^{43}_k(t,0) \right]\Bigg\},
\label{spectb}
\end{eqnarray}
with the elements of the matrix $\hat{M}$ as defined in Eq. (\ref{specta}).

\section{numerical results and dressed state analysis}
The squeezing spectra of the fluorescence fields can be obtained numerically using Eqs. (\ref{specta})
and (\ref{spectb}). In the following, we assume equal dipole moments $|\vec{\mu}_{13}| = |\vec{\mu}_{23}| = \mu$
and decay rates $\gamma_1 = \gamma_2 = \gamma$ for the upper atomic transitions. All the frequency parameters
such as decay rates, detuning, and Rabi frequencies are scaled in units of $\gamma_3$. The numerical results are
presented by considering both the presence $(p = 1)$ and absence $(p = 0)$ of quantum interference.
\begin{center}
{\bf A. Squeezing spectrum: $S_a(\omega,\theta)$}
\end{center}

We first consider spectral squeezing in the fluorescence field generated by the upper atomic transitions.
A selected quadrature component $(\theta)$ is said to exhibit spectral squeezing if the squeezing spectrum is negative,
$S(\omega,\theta) < 0$, at a certain frequency $\omega$. We analyze the squeezing spectrum $S_a(\omega,\theta)$ calculated
using Eq. (\ref{specta}) for the case of strong driving fields $(\Omega_1, \Omega_2, \Omega_3 \gg \gamma, \gamma_3)$ \cite{note1}.
In the calculations, we assume $e^{2i\omega_a r/c} = 1$ and scale the spectrum in units of $\mu^2 f(r)^2/(\pi \gamma_3)$.
The numerical results of the spectrum $S_a(\omega,\theta)$ are displayed in Fig. 2 for the in-phase $(\theta = 0)$ quadrature
component \cite{note2}. It is seen that the presence of quantum interference $(p = 1)$ enhances (three times) the spectral
squeezing $S_a(\omega,\theta) < 0$ in outer sidebands [see Fig. 2(a)].
For a suitable choice of parameters as in Fig. 3,
the squeezing can be enhanced in inner sidebands as well. Note that the amount of maximal squeezing due to quantum interference
is appreciable only when the excited atomic levels decay much slower than the middle level $\gamma_1, \gamma_2 \ll \gamma_3$

A physical understanding of the numerical results can be obtained if the squeezing spectrum (\ref{specta}) is rewritten
as a sum of two contributions, i.e., $S_a(\omega,\theta) = S_{NI}(\omega) + p S_{I}(\omega)$. Here, the terms $S_{NI}(\omega)$
and $S_{I}(\omega)$ referred to as noninterfering and interfering terms, are expressed as
\begin{eqnarray}
S_{NI}(\omega,\theta) &=& S_1(\omega) + S_2(\omega), \\
S_{I}(\omega,\theta) &=& p \left[S_{12}(\omega) + S_{21}(\omega)\right],
\end{eqnarray}
\begin{figure}[t]
   \centering
   \includegraphics[height=6cm]{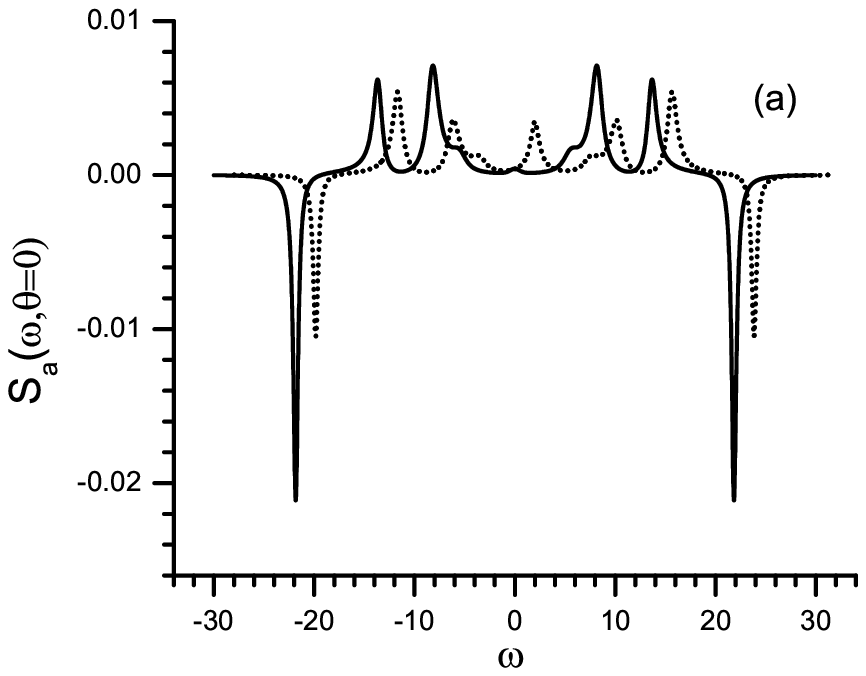}
   \includegraphics[height=6cm]{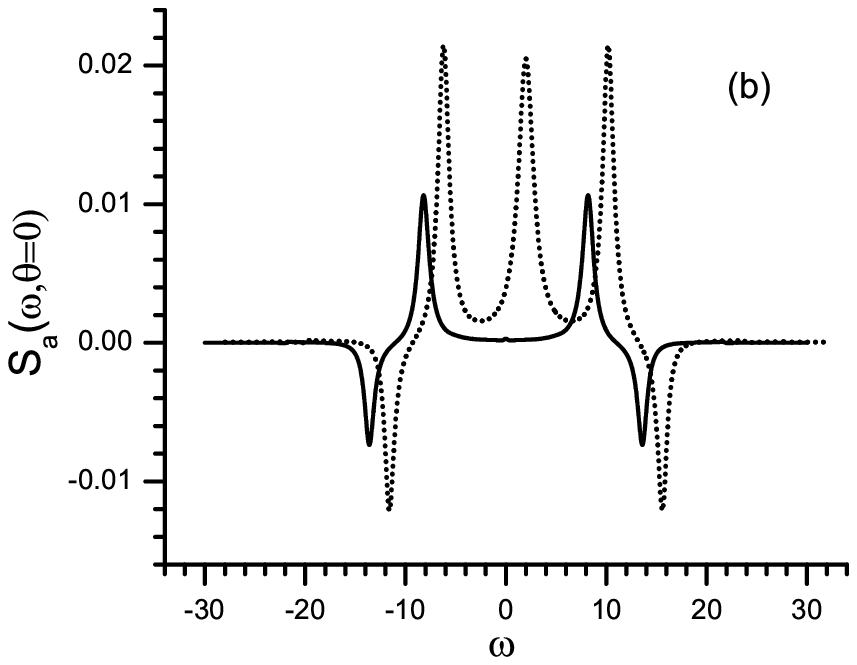}
   \caption{Squeezing spectrum $S_a(\omega,\theta)$ as a function of $\omega$ from different
   contributions: $S_1(\omega)$ [dotted curve (a)], $S_{12}(\omega)$ [solid curve (a)], $S_2(\omega)$ [dotted curve (b)],
   and $S_{21}(\omega)$ [solid curve (b)]. The parameters are those used to produce Fig. 2(a) with $p = 1$.
   For clarity, the dotted curve has been displaced by 2 units along the $\omega$-axis.} 	
\end{figure}
where
\begin{eqnarray}
S_j(\omega) = &&\frac{\mu^2 f(r)^2}{\pi} \hbox{Re} \int_{0}^{\infty} d\tau (e^{i\omega\tau} + e^{- i\omega\tau}) \nonumber \\
&& \times \left[\langle A_{3j}(t+\tau),A_{3j}(t)\rangle + \langle A_{j3}(t+\tau),A_{3j}(t)\rangle
\right], \nonumber
\end{eqnarray}
\begin{eqnarray}
S_{jk}(\omega) = && \frac{\mu^2 f(r)^2}{\pi} \hbox{Re} \int_{0}^{\infty} d\tau (e^{i\omega\tau} + e^{- i\omega\tau}) \nonumber \\
&& \times \left[\langle A_{3j}(t+\tau),A_{3k}(t)\rangle + \langle A_{j3}(t+\tau),A_{3k}(t)\rangle
\right], \nonumber \\
&&~~~~~~~~~~~~~~~~~~~~~~~~~~~~~~~~~~(j,k = 1,2).
\end{eqnarray}
The term $S_1(\omega)$ [$S_2(\omega)$] represents single transitions $|1\rangle \rightarrow |3\rangle$ [$|2\rangle \rightarrow |3\rangle$],
whereas the terms $S_{12}(\omega)$ and $S_{21}(\omega)$ correspond to vacuum mediated transitions between the excited levels $|1\rangle$
and $|2\rangle$. In Fig. 4, we plot the contributions of the different terms $S_1(\omega)$, $S_2(\omega)$, $S_{12}(\omega)$, and $S_{21}(\omega)$
as a function of frequency $\omega$ for the same parameters of Fig. 2(a). It is found that the squeezing at the outer sidebands of the spectrum
in Fig. 2(a) can be approximated as $S_a(\omega,0) \approx S_1(\omega) + p S_{12}(\omega)$.

To explore further the origin of interference induced effects, we go to the dressed state description of the atom-field interaction.
The dressed states are defined as eigenstates of the time independent Hamiltonian $H_I$ in Eq. (\ref{ham}).
In the general parametric conditions, it is difficult to find analytical solutions
to the eigenvalue problem $H_I |\Phi\rangle = \hbar \lambda |\Phi\rangle$. However, the eigenvalues $\lambda_i (i = \alpha, \beta,
\kappa, \delta)$can be obtained numerically by solving a quartic equation. The eigenstates $|\Phi_i\rangle (i = \alpha, \beta,
\kappa, \delta)$ can be expanded in terms of the bare atomic states as
\begin{equation}
|\Phi_i\rangle = a_{1i} |1\rangle + a_{2i} |2\rangle + a_{3i} |3\rangle + a_{4i}|4\rangle,
\end{equation}
where the expansion coefficients are explicitly given by
\begin{eqnarray}
a_{1i} = \frac{\lambda_i \Omega_1}{N (\lambda_i - \Delta_a - \Delta_b)}, && a_{2i} = \frac{\lambda_i \Omega_2}{N(\lambda_i + W_{12} -
\Delta_a - \Delta_b)}, \nonumber \\
a_{3i} = - \frac{\lambda_i}{N}, && a_{4i} = \frac{\Omega_3}{N} ,  \label{ecoeff}
\end{eqnarray}
with the normalization constant
\begin{eqnarray}
N &=&  \left[\Omega_3^2 + \lambda_i^2 + \frac{\lambda_i^2\Omega_1^2}{{(\lambda_i - \Delta_a - \Delta_b)}^2} \right.\nonumber \\
 &&~~~~~ \left. + \frac{\lambda_i^2\Omega_2^2}{{(\lambda_i + W_{12} - \Delta_a - \Delta_b)}^2}\right]^{\frac{1}{2}} . \nonumber
\end{eqnarray}
In order to interpret the numerical results, we consider transitions between the dressed states with the inclusion of decay processes.
The allowed transitions between the dressed states $|\Phi_i\rangle \leftrightarrow |\Phi_j\rangle (i,j = \alpha, \beta, \kappa, \delta)$
give the peaks in the fluorescence spectrum at the frequencies $\omega_{ij} = \lambda_i - \lambda_j$. Specifically,
for the parameters of Fig. 2, the eigenvalues (in units of $\gamma_3$) obtained numerically are $\lambda_{\alpha} = 20.93 $,
$\lambda_{\beta} = -0.93$, $\lambda_{\kappa} = 12.74$, and $\lambda_{\delta} = 7.26$. The squeezing found at the outer sideband in Fig. 2
can be seen as arising from the transitions $|\Phi_\alpha\rangle \leftrightarrow |\Phi_\beta\rangle$.
In the high field limit $(\Omega_1, \Omega_2, \Omega_3 \gg \gamma_1, \gamma_2, \gamma_3)$, the squeezing spectrum
$S_a(\omega,0) \approx S_1(\omega) + p S_{12}(\omega)$ can be worked out to be
\begin{widetext}
\begin{equation}
S_a(\omega_{\pm},0) = \Gamma_{\alpha\beta} \frac{(a_{3\alpha} a_{1\beta} + a_{1\alpha} a_{3\beta}) \left[(a_{3\beta} a_{1\alpha} +
p a_{3\beta} a_{2\alpha}) \rho_{\alpha \alpha}  + (a_{3\alpha} a_{1\beta} + p a_{3\alpha} a_{2\beta})  \rho_{\beta \beta} \right]}
{\Gamma_{\alpha\beta}^2 + {(\omega \mp \omega_{\alpha \beta})}^2},
\label{dressa}
\end{equation}
\end{widetext}
where the subindex + (-) stands for the positive $(\omega > 0)$ [negative $(\omega < 0)$] part of the spectrum and $\rho_{\alpha \alpha}
(\rho_{\beta \beta})$ represents the population of the dressed state $|\Phi_{\alpha}\rangle (|\Phi_{\beta}\rangle)$. The term
$\Gamma_{\alpha\beta}$ denotes the decay rate of coherence between the dressed states and is given explicitly in Appendix.
By using the numerical values of expansion coefficients (\ref{ecoeff}), the formula Eq. (\ref{dressa}) accounts very well for
the spectral squeezing displayed in Fig. 2.

\begin{figure}[b]
   \centering
   \includegraphics[height=6cm]{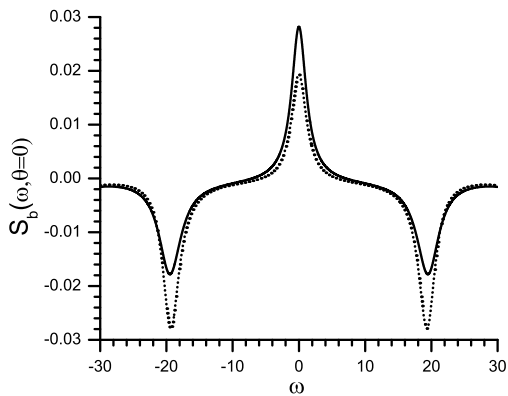}
   \caption{Squeezing spectrum $S_b(\omega,\theta)$ as a function of $\omega$ for the
   parameters $\theta = 0$, $\gamma_3 = 1$, $W_{12} = 10$, $\Delta_a = \Delta_b = 20$, $\Omega_1 = \Omega_2 =
   \Omega_3 = 6$, and $\gamma_1 = \gamma_2 = 3.$ The solid (dotted) curves are for $p = 1$ $(p = 0)$.} 	
\end{figure}
\begin{center}
{\bf B. Squeezing spectrum: $S_b(\omega,\theta)$}
\end{center}

We next consider the spectrum of fluorescence field generated by the lower atomic transitions. The squeezing spectrum calculated using
Eq. (\ref{spectb}) is displayed in Fig. 5 for the in-phase $(\theta = 0)$ quadrature \cite{note2}. The spectrum is scaled in units of
${|\vec{\mu}_{34}|}^2 g(r)^2/(\pi \gamma_3)$. We also assume $e^{2i\omega_b r/c} = 1$ and consider the case of fast decaying upper
levels $(\gamma_1, \gamma_2 \gg \gamma_3)$ of the atom. The effect of quantum interference $(p = 1)$ is now seen to reduce the squeezing in
spectral components as evident from Fig. 5. For slow decay of the upper levels $(\gamma_1, \gamma_2 \preceq \gamma_3)$, the reduction in
squeezing is not appreciable (graph not shown).
\begin{figure}[t]
   \centering
   \includegraphics[height=6cm]{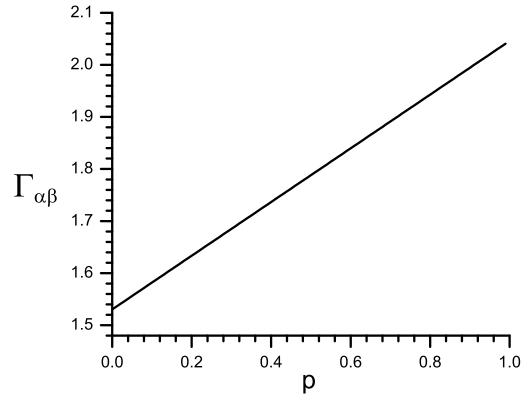}
   \caption{The decay rate $\Gamma_{\alpha \beta}$ as a function of the interference parameter
   $p$. The other parameters for the calculation are the same as in Fig. 5.} 	
\end{figure}
In order to understand this result, the spectrum is further analyzed in the dressed state
picture. The contribution to the squeezing spectrum (\ref{spectb}) originating from the dressed-state transitions $|\Phi_{\alpha}\rangle
\leftrightarrow |\Phi_{\beta}\rangle$ can be given as
\begin{eqnarray}
S_b(\omega_{\pm},0) &=& \Gamma_{\alpha\beta}\frac{(a_{3\alpha} a_{4\beta} + a_{4\alpha} a_{3\beta})}{\Gamma_{\alpha\beta}^2 +
{(\omega \mp \omega_{\alpha \beta})}^2} \nonumber \\
&& \times \left[a_{3\alpha} a_{4\beta} \rho_{\alpha \alpha} + a_{4\alpha} a_{3\beta} \rho_{\beta \beta}\right],
\label{dressb}
\end{eqnarray}
where the different terms have their meanings as in Eq. (\ref{dressa}). For the parameters of Fig. 5, the numerical values of eigenvalues
(in units of $\gamma_3$) obtained are $\lambda_{\alpha} = 17.55 $, $\lambda_{\beta} = -1.82$, $\lambda_{\kappa} = 41.99$, and
$\lambda_{\delta} = 32.28$. The spectrum Eq. (\ref{dressb})is a pair of Lorentzians centered at the frequencies $\pm \omega_{\alpha \beta}
= \pm (\lambda_{\alpha} - \lambda_{\beta})$. Since the numerators of the Lorentzians are negative, the graph (Fig. 5) has negative peaks of
height proportional to the inverse of decay rate $\Gamma_{\alpha\beta}$.The reduction in squeezing observed in Fig. 5 may be traced to the
increase in the decay rate $\Gamma_{\alpha\beta}$ of the dressed atomic transition. To this end we study the variation of the decay rate
$\Gamma_{\alpha\beta}$ as a function of the interference parameter $p$. As seen in Fig. 6, the decay rate attains a maximum in the presence
of full quantum interference $(p=1)$, thereby reducing the spectral squeezing observed in Fig. 5.

Finally, we note that the spectrum $S_b(\omega,\theta)$ exhibits pure two-level squeezing for weak applied fields
$(\Omega_1, \Omega_2, \Omega_3 \ll \gamma_1, \gamma_2, \gamma_3)$ independent of the quantum interferences. The spectrum in this case is centered
around the laser frequency $(\omega = 0)$ and the maximum squeezing is obtained for the out-of-phase quadrature $(\theta = \pi/2)$
similar to two-level atoms \cite{collet}.

\section{summary}
In this paper, we investigated the squeezing spectrum of the resonance fluorescence from a driven Y-type atom
when the presence of interference in spontaneous decay channels is important. In particular, we considered the atom
to be driven by two coherent fields and examined the squeezing spectrum of fluorescence radiation from both the upper-
and lower transitions in the atoms. It was shown that the decay-induced interference enhances squeezing in the
spectrum of upper transitions for off-resonance and strong driving fields. A detailed analysis using dressed-states
was presented to bring out the role of interferences. Further, the interference was also shown to degrade the spectral
squeezing if the fluorescence is detected on the lower atomic transitions. This has been explained as due to a fast
decay of the dressed state in the atom.

\appendix*
\section{}
In the secular approximation, the time evolution of the coherence $\rho_{\alpha \beta}(t)$ between the dressed states $|\Phi_{\alpha}\rangle$ and $|\Phi_{\beta}\rangle$ obeys
\begin{equation}
\frac{d\rho_{\alpha \beta}}{dt} = -(\Gamma_{\alpha \beta} + i \omega_{\alpha \beta}) \rho_{\alpha \beta,}
\end{equation}
with $\omega_{\alpha \beta} = \lambda_{\alpha} - \lambda_{\beta}$ obtained by diagonalizing the Hamiltonian (\ref{ham}). The decay rate
$\Gamma_{\alpha \beta}$ is given by
\begin{equation}
\Gamma_{\alpha \beta} = \Gamma_1 \gamma_1 + \Gamma_2 \gamma_2 + \Gamma_3 \gamma_3 + \Gamma^{\prime} p\sqrt{\gamma_1 \gamma_2},
\end{equation}
where
\[
\Gamma_1 = a_{1\alpha}^2 + a_{1\beta}^2 - 2 a_{1\alpha} a_{1\beta} a_{3\alpha} a_{3\beta},
\]
\[
\Gamma_2 = a_{2\alpha}^2 + a_{2\beta}^2 - 2  a_{2\alpha} a_{2\beta} a_{3\alpha} a_{3\beta},
\]
\[
\Gamma_3 = a_{3\alpha}^2 + a_{3\beta}^2 - 2 a_{3\alpha} a_{3\beta} a_{4\alpha} a_{4\beta},
\]
\[
\Gamma^{\prime} = 2 a_{1\alpha} a_{2\alpha} + 2 a_{1\beta} a_{2\beta} - 2 a_{3\alpha} a_{3\beta} (a_{1\alpha} a_{2\beta} +a_{1\beta} a_{2\alpha}).
\]


\begin{thebibliography}{99}
\bibitem{note1} For weak driving fields $(\Omega_1, \Omega_2, \Omega_3 \ll \gamma_1, \gamma_2, \gamma_3)$,
the spectrum $S_a(\omega,\theta)$ generated by the upper transitions does not show squeezing in any quadrature in the
absence and in the presence of quantum interference.
\bibitem{note2} The spectral squeezing appears in out-of-phase $(\theta = \pi/2)$ quadrature and
more generally other phase quadratures $(\theta \neq 0)$ as well. However, the in-phase quadrature
exhibits more squeezing than the out-of-phase quadrature.
\end{thebibliography}
\end{document}